# A Merger Origin for X-structures in S0 Galaxies

J. Christopher Mihos, Ian R. Walker, Lars Hernquist,[1]
*Board of Studies in Astronomy and Astrophysics*
*University of California, Santa Cruz, CA 95064*
*hos@lick.ucsc.edu, iwalker@lick.ucsc.edu, lars@lick.ucsc.edu*

Claudia Mendes de Oliveira
*European Southern Observatory*
*Casilla 19001, Santiago, 19, Chile*
*and*
*Instituto Astronômico e Geofísico*
*USP, C.P. 9638,*
*São Paulo, 01065-970, Brazil*
*coliveir@eso.org*

and

Michael Bolte
*Lick Observatory*
*University of California, Santa Cruz, CA 95064*
*bolte@lick.ucsc.edu*

## ABSTRACT

Using numerical simulation, we study the response of a disk galaxy to a merger involving a low-mass satellite companion. During a prograde satellite accretion, the disk galaxy forms a strong bar in response to the perturbation of the companion. After the accretion event is over, the bar buckles vertically due to a bending instability, sending disk material well out of the disk plane. The material forms into an X-shaped feature when seen edge-on, morphologically similar to X-structures observed in several S0/Sa galaxies. The mechanism described here unites previously suggested accretion and bar instability hypotheses for forming X-structures into a self-consistent scenario for merger-induced formation and evolution of S0 galaxies. To complement these models, we also present observations of the peculiar S0 galaxy Hickson 87a. The thick disk, isophotal warping, and strong X-structure described in the merger model are all evident in Hickson 87a, suggesting this galaxy may be an excellent example of such merger-induced galaxy evolution.

*Subject headings:* galaxies:evolution – galaxies:interactions – galaxies:peculiar – galaxies:structure

## 1. Introduction

Morphological studies of edge-on disk galaxies have revealed that a significant fraction of them possess peculiar boxy or peanut-shaped isophotes. The exact fraction of galaxies harboring such features is still uncertain, being subject to extreme selection effects, but estimates range from a few percent (Jarvis 1986) to as many as 20% (Shaw 1987). In fact, recent COBE observations of the bulge of the Milky Way suggest that our own Galaxy has a peanut-shaped bulge (Weiland et al. 1994). An interesting subset of these galaxies are those whose bulges contain sharp X-structures, e.g., NGC 4845, NGC 1381, IC 4767 (Whitmore & Bell 1988), IC 3370 (Jarvis 1987), and AM 1025-401 (Arp & Madore 1987). These galaxies are all early-type S0/Sa disk galaxies, which suggests some commonality between processes driving the formation of X-structures and those leading to the formation of S0 galaxies.

The origin of these peculiar bulges has remained somewhat of a mystery. Several mechanisms have been put forward, each with accompanying caveats. Binney & Petrou (1985) suggested that these structures may result from material accreted from an infalling satellite companion, an idea echoed later by Whitmore & Bell (1988) in their study of IC 4767. However, the luminosity in the X in IC 4767 is comparable to the luminosity of the disk component; such a massive accretion event would surely have disrupted the stellar disk (e.g., Barnes 1988, 1992; Hernquist 1992, 1993a). An alternative mechanism for forming peculiar bulges involves the buckling of a strong bar in the disk (Combes & Sanders 1981; Combes et al. 1990; Raha et al. 1991). In this scenario the material forming the peculiar structures comes from the disk itself, so there is no inconsistency with the large fraction of light found in the X of IC 4767. Furthermore, the observed stellar kinematics in several boxy and peanut-shaped bulge galaxies are consistent with the velocity fields in models of barred galaxies (Bettoni & Galletta 1994; Kuijken & Merrifield 1995). While the models of Combes et al. and Raha et al. were able to produce features morphologically similar to those observed in many boxy-bulged galaxies, the initial conditions for the disk galaxy models employed in those studies were unstable against bar formation, leaving unanswered the question of how disk galaxies could evolve into this state initially. Fisher, Illingworth, & Franx (1994) suggested that galaxy interactions could drive bar formation in dynamically stable disk galaxies, leading to the formation of a boxy bulge. An additional mechanism for forming peculiar bulge-like features in disk galaxies involves bending instabilities in counterrotating stellar disks (Sellwood & Merritt 1994); again, however, the question of initial conditions remains unanswered.

In this *Letter*, we present an evolutionary scenario which unites in a self-consistent manner the accretion and bar hypotheses for the formation of X-structures in S0 galaxies. Using numerical simulation, we show that a merger between a disk galaxy and small satellite companion can excite a strong bar mode in a galactic disk. The bending of this long-lived bar populates stellar orbits which carry disk material high above the disk plane and results in the formation of the X. These numerical models were originally run to study the evolution of satellite accretion events, and it was only when the models were analyzed that we noticed a close morphological match to the peculiar S0 galaxy Hickson 87a (Hickson 1993). Subsequent optical and near infrared imaging of this galaxy confirms quantitatively the similarity between Hickson 87a and our numerical models. We conclude with a brief discussion of our results in the context of current theories of galactic evolution.

## 2. Merger Model

As part of a larger study of the effects of satellite mergers on the structure of the primary galaxy (Walker, Mihos, & Hernquist 1995), we model the purely stellar-dynamical evolution of a low mass companion galaxy being accreted by a larger disk galaxy. In this encounter, the primary galaxy is comprised of a spherical, roughly isothermal dark halo and exponential stellar disk, with a halo to disk mass ratio of 5.8:1 (see Hernquist 1993b for details). The satellite companion is constructed using the Hernquist (1990) potential-density pair for spherical galaxies, and has a total mass of 0.1, or 10% of the disk mass. We use a total of 500,000 particles to model the system, 225,000 in each of the galactic disk and halo and 50,000 in the satellite companion. The system is evolved using a fully self-consistent treecode (Hernquist 1987), conserving energy and angular momentum to better than 1%.

Figure 1 shows the dynamical evolution of the merging disk/satellite system. The satellite galaxy is placed on an initially circular orbit at a radius of six disk scale lengths. The orbit of the satellite is largely prograde, inclined by 30° to the disk plane (the effect of differing orbital geometries is discussed in §4). As the satellite orbits at large radius, it warps the outer disk before losing its $z$ motion and falling into the disk plane. Once the satellite settles into the disk plane, enhanced dynamical friction causes the satellite to rapidly shed angular momentum and spiral into the center of the disk. As it does so, the satellite significantly heats the stellar disk and drives a strong $m = 2$ bar mode which lasts until the end of the simulation, $\sim 1.5$ Gyr after the satellite has reached the center.

As the post-merger galaxy evolves, the strength of the bar (as measured by the strength of the $m = 2$ Fourier mode) is relatively constant over about six rotation periods ($\sim 1$ Gyr). With time the bar flexes, or buckles (Figure 2), in much the same way as in the models of Raha et al. (1991), as a result of the well-known bending instability in galactic systems (Toomre 1967; Merritt & Hernquist 1991; Hernquist, Heyl, & Spergel 1993; Merritt & Sellwood 1994). This buckling of the stellar bar sends material well out of the disk plane, forming a prominent X-shaped feature from stars which originally resided in the stellar disk. It is important to emphasize that while the perturbation of the infalling satellite drives the growth of the stellar bar, the X-structure in the disk is *not* comprised largely of satellite material. Unlike previously suggested accretion mechanisms (Binney & Petrou 1985; Whitmore & Bell 1988), which did not account for the early settling of the satellite into the disk plane, our model has satellite material deposited not on inclined orbits at small radii, but directly into the disk plane where it spatially mixes with the original disk material. Because the X forms from *disk* material, the fact that the disk and X in IC 4767 have comparable luminosities (Whitmore & Bell 1988) does not imply that IC 4767 accreted a comparable fraction of mass.

To illustrate this X-feature, Figure 3 shows the model at a time 1.5 Gyr after the merger is complete. To produce this figure, we project the three-dimensional model onto a "sky" plane to create a surface density map. Assuming that the disk and satellite have a similar mass-to-light ratios, and ignoring dust obscuration, we can treat this map as an observed surface brightness map – an "image" analogous to a broad band CCD image. For the figure, we choose to view the galaxy edge-on with the major axis of the bar in the plane of the sky. By contrast, if we choose to view along the bar axis, the X-structure is much less noticeable, and the galaxy appears to have much rounder central isophotes (see also Combes et al. 1990).

Several features are immediately obvious upon inspection of Figure 3. The disk itself has been considerably thickened, both by the direct effects of the infalling satellite (e.g., Quinn, Hernquist, & Fullager 1993; Walker et al. 1995) and by the subsequent evolution of the induced bar. The thickness of the disk (as measured most simply by $\langle |z| \rangle$ of the disk particles) has effectively doubled and is now reminiscent of the thick stellar disks in early-type S0 galaxies (e.g., Burstein 1979). Furthermore, the outer isophotes display significant warping caused by the interaction between the disk and satellite early in the encounter. At higher surface brightness levels, the remnant of the accreted satellite can be seen as a dense clump at the center of the accreting disk.

Most interesting for our purposes, however, is the X-shaped feature seen at low isophotal intensity levels. As shown in these images, this feature is very reminiscent of similar structures in IC 4767 (Whitmore & Bell 1988) and AM 1025-401 (Arp & Madore 1987). Figure 3c shows an enhanced image highlighting this X-structure. To create this image, the original image is smoothed with a 20-pixel boxcar window, comparable in size to the thickness of the stellar disk. This boxcar-smoothed image is then subtracted from the original image to create the image shown in Figure 3c. This process, known as "unsharp masking," has the effect of enhancing structure on size scales comparable to or smaller than the smoothing window (Malin 1988). In this image, the X feature can be seen extending far above and below the disk plane, and appears distinct from classical peanut-shaped bulges, which typically have much rounder isophotes (see, for example, Jarvis 1987).

## 3. Comparison with Hickson 87a

CCD observations of Hickson 87a (H87a) were first made as part of a program to obtain photometric and spectroscopic parameters for all galaxies in Hickson's compact groups (Hickson 1993). This galaxy is classified by the RC3 (de Vaucouleurs et al. 1991) as a peculiar S0 (although it is classified as an Sbc by Hickson) with a velocity of 8694 km s$^{-1}$ and an absolute B magnitude of $-19.78 \times 5 \log h_{100}$. As indicated by its B–R color of ∼ 1.86 (Hickson 1993), H87a is one of the reddest spiral galaxies in Hickson's compact group sample due to a strong, complex dust lane present in its disk. H87a has three close neighbors on the sky within a projected radius of $30h^{-1}$ kpc, with velocities 8972, 8920 and 10200 km/s, the last being a background galaxy projected on the group.

Figure 4 shows a 120-second R-band exposure of Hickson 87a taken in sub-arcsecond seeing with the NTT and the EMMI camera on 5-Oct-1994. In the deep stretch (Fig 4a), the similarity to the final state of the merger model is obvious: the thick disk, the warped outer isophotes, and the "horns" of the X projecting out of the disk plane are all apparent. A shallower stretch (Fig 4b) reveals the complexity of the dust lane in the disk. On the western side, the dust lane is narrow and linear, while on the eastern side it shows a split, twisted morphology, suggesting severe warping of the disk, perhaps brought on by a merger event. The high surface brightness central core – partially obscured by the dust lane in the disk – also is striking in its similarity to the satellite remnant in the merger model. To highlight the X-structure in more detail, we construct an unsharp masked image in an identical manner to that used for the merger model. This image, shown in Figure 4c, again shows the strong X-feature extending from the inner regions of the galaxy to well above and below the disk plane.

As a more quantitative measure of the similarities between the model and H87a, we also perform an isophotal analysis to compare the strength of the two X components. Because of the thickness and warping of both the model disk and the disk in H87a, we choose not to attempt to fit standard edge-on disk isophotes to the systems. Rather, we take the more simplistic approach of fitting elliptical isophotes and examining the isophotal deviations from pure ellipses. Of particular interest is the strength of the $cos(4\theta)$ term in the Fourier expansion of these deviations (hereafter called $c_4$), which describes "boxy" or "disky" isophotes (e.g., Lauer 1985; Bender & Mollenhoff 1987). To avoid complications in these fits due to the strong dust lane in H87a, we use a 10 × 60-second K band exposure obtained on the ESO 2.2m with the IRAC2 camera on 6-Oct-1994 rather than the R band image shown in Figure 4.

Figure 5 shows the results of this isophote fitting. Because the pixel scales for the model and H87a do not necessarily match, a radial scaling between the two images must be performed to allow a comparison of $c_4$ as a function of radius. To choose this scaling in a relatively unbiased manner, we alter the scaling of the model image pixels until the light profile of the galaxy model and H87a fall off in roughly the same fashion (see Figure 5a). Using this scaling, we then compare the strength of $c_4$ as a function of radius in Figure 5b. In both the model and H87a, the isophotes show a tendency to be rather boxy, at the 2% level, over similar ranges in radius. This more quantitative comparison of isophote morphology re-emphasizes the strong similarity between the model and Hickson 87a.

## 4. Discussion

The mechanism presented here unites into a self-consistent picture the accretion and bar hypotheses for the formation of X-structures in S0 galaxies. Dynamical heating due to the merging satellite accounts for the thickened, warped galactic disk, while the formation and subsequent evolution of the stellar bar gives rise to the X-structure seen in an edge-on view. While the mechanism for forming the X-structure is very similar to that described by Raha et al. (1991), the model shown here employs dynamically stable galaxy models as the progenitor galaxies and describes the formation of X galaxies self-consistently in the context of merger-induced galaxy evolution.

Galaxy mergers have long been suspected of driving significant galaxy evolution. While much attention has been paid to the elliptical-like remnants of "major" mergers of comparable-mass disk galaxies (e.g., Toomre 1977; Schweizer 1982), it has also been proposed that the presumably more common "minor" satellite mergers studied here may also transform the Hubble type of the primary disk (Schweizer & Seitzer 1988; Hernquist 1989, 1991; Schweizer 1992; Mihos & Hernquist 1994). If minor mergers do transform late-type disk galaxies into earlier S0/Sa galaxies, these galaxies may show signatures of the merging process such as X-structures and thickened disks which can be used to constrain the fraction of galaxies so affected. We see all these morphological signatures in the peculiar S0 galaxy Hickson 87a, which strongly suggests that this galaxy is a prime example of this type of merger-induced galaxy evolution.

Do all satellite accretions produce X-structures in the primary disk galaxy? The merging sequence which gives rise to the X-structure requires both the formation of a strong bar and the subsequent long-term (i.e. few Gyr) survival of the bar. Not all satellite merger scenarios satisfy both these criteria. While prograde encounters are very effective at exciting bars, polar or retrograde encounters are less so, due to the weaker coupling between the orbital motion of the companion and the rotation of the primary disk (Walker et al. 1995). Furthermore, the presence of a dense central bulge in the disk can act to dynamically stabilize the disk against strong bar formation even in prograde encounters (Mihos & Hernquist 1994, Hernquist & Mihos 1995). As a result, only a fraction of minor mergers may give rise to conditions necessary for the formation of X-structures. Even in interactions which do give rise to bar formation, the bars may not survive long enough to drive the formation of an X-structure. In particular, the evolution of any gaseous component in the disk may have serious effects on the longevity of the bar. Models of satellite mergers which include hydrodynamics show that such mergers can drive strong gaseous inflows in the disk (Hernquist & Mihos 1995), depending on the internal structure of the disk galaxy. The resulting central concentration of gas mass acts to destabilize the stellar orbits which populate the bar, causing dissolution of the bar over a few dynamical timescales (Hasan & Norman 1990; Hasan, Pfenniger, & Norman 1993). Accordingly, the formation of X-structures may be strongly curtailed in gas-rich disks, where gaseous inflows may act to disrupt any induced stellar bars. Because of these many complications to the merger scenario described in §2, clearly only a subset of minor mergers will result in the formation of X-structures in the primary disk galaxy.

Because X-structures can form only in a fraction of minor mergers, the relative rarity of these structures in disk galaxies is not inconsistent with merger-induced evolution of galactic disks. Unfortunately, X-structures will be difficult to use as a "smoking gun" signature of accretion events because of the special conditions necessary for their formation and the special viewing geometry necessary for their detection. To reveal any X-structure, the remnant disk galaxy must be observed edge-on, with the bar major axis near the plane of the sky. Nevertheless, the fact that Hickson 87a matches so closely the many peculiar features of the merger model – the warped, thickened disk, the peaked light profile in the nucleus, and the prominent X-structure – suggests that at least some S0 galaxies do form in such a manner.

Finally, because the X-structure in our model forms from original disk material, X-structures formed in this manner should corotate with the surrounding stellar disk. By contrast, in scenarios where X-structures form from material accreted on inclined orbits, the accreted stars forming the X can either corotate or counterrotate with respect to the stellar disk. Kinematic studies of X-structures should therefore prove useful in further constraining different scenarios for the formation and evolution of S0 galaxies.

We thank Alar Toomre, David Fisher, and Duncan Forbes for helpful discussions. This work was supported in part by the Pittsburgh Supercomputing Center, the San Diego Supercomputing Center, the Alfred P. Sloan Foundation, NASA Theory Grant NAGW–2422, the NSF under Grants AST 90–18526 and ASC 93–18185 and the Presidential Faculty Fellows Program.

## REFERENCES

Arp, H.C., & Madore, B.F. 1987, in A Catalogue of Southern Peculiar Galaxies and Associations (Cambridge: Cambridge University Press)

Barnes, J.E. 1988, ApJ, 331, 699

Barnes, J.E. 1992, ApJ, 393, 484

Bender, R. & Mollenhoff, C. 1987, A&A, 177, 71

Bettoni, D. & Galletta, G. 1994, A&A, 281, 1

Binney, J., & Petrou, M. 1985, MNRAS, 214, 449

Burstein, D. 1979, ApJ, 234, 829

Combes, F., Debbasch, F., Friedli, D., & Pfenniger, D. 1990, A&A, 233, 82

Combes, F. & Sanders, R.H. 1981, A&A, 96, 164

de Vaucouleurs, G. et al. 1991, Third Reference Catalogue of Bright Galaxies, (New York : Springer-Verlag) (RC3)

Fisher, D., Illingworth, G., & Franx, M. 1994, AJ, 107, 160

Hasan, H. & Norman, C. 1990, ApJ, 361, 69

Hasan, H., Pfenniger, D., & Norman, C. 1993, ApJ, 409, 91

Hernquist, L. 1987, ApJS, 64, 715

Hernquist, L. 1989, Nature, 340, 687

Hernquist, L. 1990, ApJ, 356, 359

Hernquist, L. 1991, Intl. J. Supercomput. Appl., 5, 71

Hernquist, L. 1992, ApJ, 400, 460

Hernquist, L. 1993a, ApJ, 409, 548

Hernquist, L. 1993b, ApJS, 86, 389

Hernquist, L., Heyl, J.S., & Spergel, D.N. 1993, ApJ, 416, L9

Hernquist, L. & Mihos, J.C. 1995, ApJ, in press

Hickson, P. 1993, Astrophysical Letters and Communications, 29, 1

Jarvis, B.J. 1986, AJ, 91, 65

Jarvis, B.J. 1987, in Structure and Dynamics of Elliptical Galaxies, ed. T. de Zeeuw (Dordrecht: Reidel) 411

Kuijken, K. & Merrifield, M.R. 1995, ApJ, in press

Lauer, T..R. 1985, MNRAS, 216, 429

Malin, D. 1988, New Scientist, 120, 23

Merritt, D., & Hernquist, L. 1991, ApJ, 376, 439

Merritt, D., & Sellwood , J.A. 1994, ApJ, 425, 551

Mihos, J.C., & Hernquist, L. 1994, ApJ, 425, L13

Quinn, P.J., Hernquist, L., & Fullager, D.P. 1993, ApJ, 403, 74

Raha, N., Sellwood, J.A., James, R.A., & Kahn, F.D. 1991, Nature, 352, 411

Schweizer, F. 1982, ApJ, 252, 455

Schweizer, F. 1992, in Physics of Nearby Galaxies: Nature or Nurture?, ed. T.X. Thuan, C. Balkowski, & J. Tran Thanh Van (Gif-sur-Yvette: Editions Frontières)

Schweizer, F., & Seitzer, P. 1988, ApJ, 328, 88

Sellwood , J.A. & Merritt, D. 1994, ApJ, 425, 530

Shaw, M.A. 1987, MNRAS, 229, 691

Toomre, A. 1967 in Notes from the Geophysical Fluid Dynamics Summer Program (Woods Hole: Woods Hole Oceanographic Institution), 111

Toomre, A. 1977 in The Evolution of Galaxies and Stellar Populations, ed. B. Tinsley & R. Larson (New Haven: Yale Univ. Press), 401

Walker, I., Mihos, J.C., & Hernquist, L. 1995, in preparation

Weiland, J.L. et al. 1994, ApJ, 425, L81

Whitmore, B.C., & Bell, M. 1988, ApJ, 324, 741

Fig. 1.— Evolution of the satellite accretion event. a) Top view, b) Side view. Each frame measures 20 disk scale lengths on a side. Elapsed time is shown in each frame; unit time is 13 million years.

Fig. 2.— A subset of disk particles in the stellar bar which end up in the X-structure, viewed 125 Myr after the merger is complete. Note the strong bending of the bar, and its saddle-shaped appearance.

Fig. 3.— Mosaic of images showing structure in the satellite merger model. The model galaxy is shown 1.5 Gyr after the merger is complete. From left to right the images are a) a deep stretch showing low surface brightness features, b) a shallower stretch showing high surface brightness features, and c) an unsharp-masked image highlighting the X-structure (see text for description).

Fig. 4.— Mosaic of images, analogous to Figure 3, showing structure in Hickson 87a.

Fig. 5.— A comparison of the isophotal properties of the merger model and Hickson 87a. Top: Mean isophotal intensity vs. semi-major axis length (a). The radial scaling of the merger model has been chosen to provide the best match between the profiles. Bottom: $c_4$, the contribution of the $\cos(4\theta)$ term to the deviations of the isophotes from their best fitting ellipses, vs. semi-major axis length.